\documentclass[twoside]{mhd}
\usepackage{graphicx}
\usepackage{bm}
\mhdhead{40}{1}{1}

\title{\uppercase{ALFV\'EN WAVE EXPERIMENTS WITH LIQUID RUBIDIUM IN A
PULSED MAGNETIC FIELD}}

\author{Th.~Gundrum\inst{1}, J.~Forbriger\inst{1}, 
Th.~Herrmannsd\"orfer\inst{1},\\ G.~Mamatsashvili\inst{1}, S. Schnauck\inst{1,2}, 
F.~Stefani\inst{1}, J. ~Wosnitza\inst{1} 
}

\institute{Helmholtz-Zentrum Dresden -- Rossendorf, Bautzner Landstr. 400, 
01328 Dresden, Germany 
\and
Anton Pannekoek Institute for Astronomy, University of Amsterdam, 
Postbus 94249, 1090 GE Amsterdam, The Netherlands}

\begin{document}
\maketitle
% ------------------------------ Abstract --------------------------------

\noindent
\textbf{Abstract:}
Magnetic fields are key ingredients for heating the solar corona 
to temperatures of several million Kelvin. A particularly important 
region with respect to this is the so-called
magnetic canopy below the corona, where sound and Alfv\'en waves have 
roughly the same speed and can, therefore, easily transform into each 
other. We present the results of an Alfv\'en-wave experiment with liquid 
rubidium carried out in a pulsed field of up to 63\,T. At the 
critical point of 54\,T, where the 
speeds of Alfv\'en waves and sound coincide, a new 4\,kHz
signal appears in addition to the externally excited 8\,kHz 
torsional wave. This emergence of an Alfv\'en wave with a doubled period is in 
agreement with the theoretical predictions of a parametric resonance 
between the two wave types. We also present preliminary results from 
numerical simulations of Alfv\'en and magneto-sonic waves 
using a compressible MHD code.

%\begin{keywords} Tayler instability, Tayler-Spruit dynamo, helicity oscillations \end{keywords}
% --------------------------- Introduction--------------------------------

\section{Introduction}
\label{sec:intro}
Shortly after their theoretical prediction in 1942 \cite{Alfven1942}, the
existence of Alfv\'en waves was confirmed in liquid-metal experiments by Lundquist 
\cite{Lundquist1949} and Lehnert \cite{Lehnert1954}. Later, they were studied in 
much detail in large-scale plasma experiments, a
comprehensive summary of which can be found in the paper by Gekelman \cite{Gekelman1999}. Nowadays,
Alfv\'en waves are known to play a key role in many phenomena in astrophysical and 
fusion-related plasmas \cite{Cramer}. They are, in particular, one of the main 
candidates to explain the heating of
the solar corona \cite{Tomczyk2007,Grant2018,Srivastava2017} to temperatures of several 
million Kelvin. A particularly important
region in this matter is the so-called magnetic canopy below the corona (see Fig. 1), where
sound and Alfv\'en waves have roughly the same speed and can, therefore, easily transform
into each other \cite{Zaqarashvili2006}. However, this ``magic point'' 
has remained inaccessible to
experiments until recently: while in plasma experiments the Alfv\'en speed is typically much
higher than the speed of sound, in all previous liquid-metal experiments it has been
significantly lower.

\begin{figure}[t]
  \centering
  \includegraphics[width=0.5\textwidth]{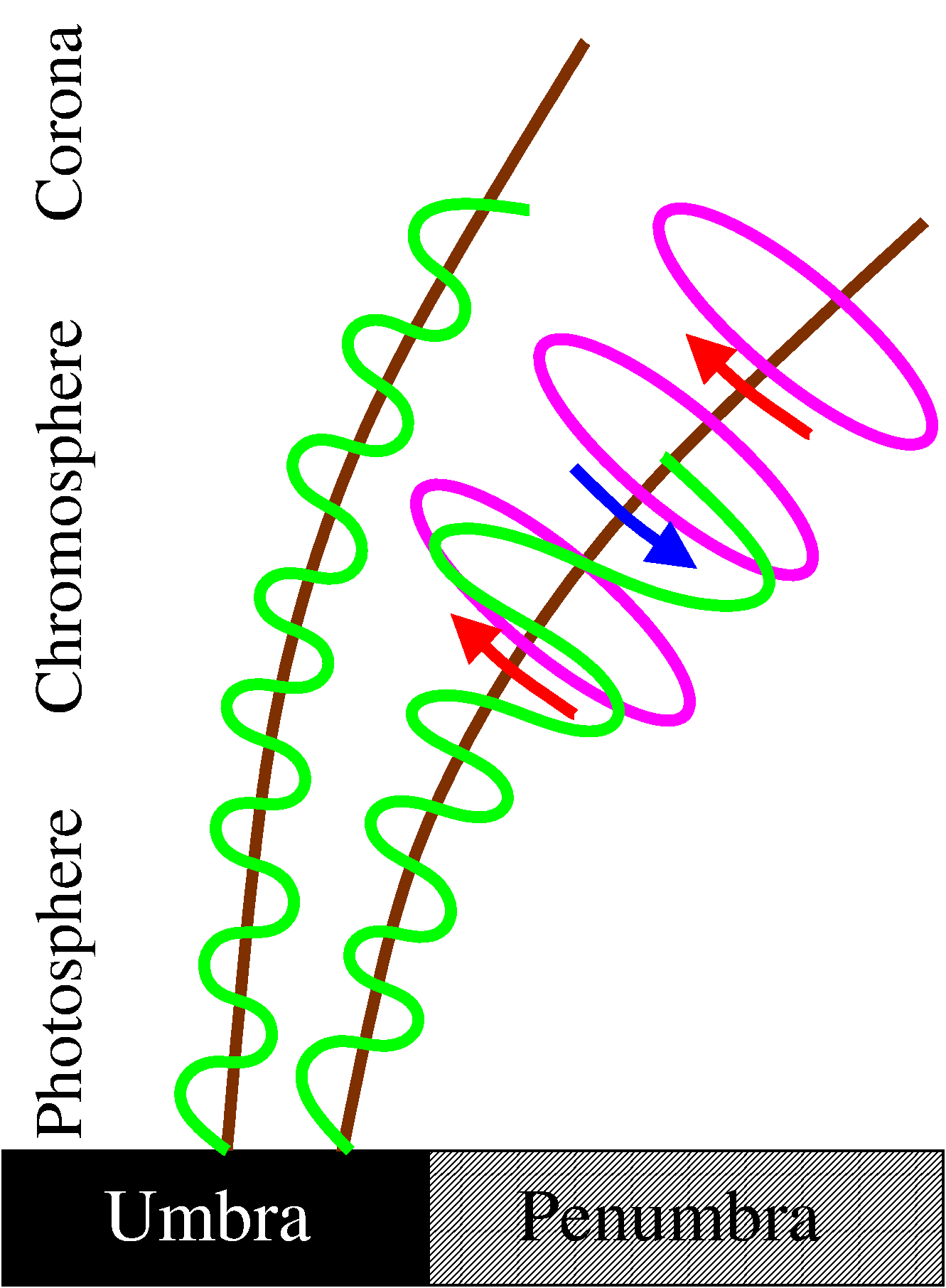}
  \caption{Illustration of the transformation of magneto-sonic waves (green) into 
  period-doubled torsional Alfv\'en waves (purple) in the solar chromosphere where 
  $v_a \approx c_s$ (after
\cite{Grant2018}). Such torsional waves, travelling along magnetic field lines (brown),
 have recently been measured in the
solar chromosphere \cite{Srivastava2017}.
  }
  \label{fig:1}
\end{figure}

Recent developments in reaching pulsed magnetic fields above 50\,Tesla have 
revealed new prospects for Alfv\'en-wave experiments with liquid metals. The main idea of our
experiment \cite{Stefani2021} was, therefore, to utilize high pulsed magnetic fields to allow 
Alfv\'en waves whose propagation speed increases proportionally to the 
magnetic field to cross 
the sound speed of $c_s = 1260$\,m/s at the ``magic point'' of 54\,T, when using liquid
rubidium. By injecting an alternating current at the bottom of the rubidium container and
exposing it to the pulsed magnetic field of up to 63\,T (Fig. 2), it became possible to generate
Alfv\'en waves in the melt, whose upward motion was measured at the expected speed,
similarly to previous experiments \cite{Alboussiere2011}. In that paper, 
we described that, while all measurements up to the
``magic point'' of 54 T were dominated by the 8\,kHz signal of the injected
AC current, exactly at this point a new 4 kHz signal appeared. This sudden emergence of
period doubling was in perfect agreement with the theoretical predictions of a parametric
resonance as described by Zaqarashvili and Roberts \cite{Zaqarashvili2006}. 
According to their theory, the Alfv\'en wave
drives the sound wave through the ponderomotive force, while the sound wave returns
energy back to the Alfv\'en wave through swing excitation.
Here, we summarize the main experimental 
outcomes of  \cite{Stefani2021}, and complement them with first numerical
results on the interaction of Alfv\'en- and magneto-sonic waves that were obtained with 
the compressible MHD-code PLUTO \cite{PLUTO}.

\section{The experiment}

The experiments were carried out within a long-pulse
coil \cite{Wosnitza2007} of the Dresden High Magnetic Field Laboratory (HLD) at 
Helmholtz-Zentrum Dresden-Rossendorf (HZDR). Situated in its central bore of 24 mm diameter, the (warm) liquid rubidium
experiment was shielded thermally from the liquid-nitrogen cooled copper coil by a
Dewar wall. The rubidium-filled stainless steel container (Fig. 2a), with inner diameter 10 mm
and 60 mm height, was embedded into a holder for the pick-up and compensation coils (Fig. 2b).
During the experiment, a controlled current source (Agilent 33220A and Rohrer PFL-2250-
28-UDC415-DC375) delivered a very stable sinusoidal cw current with constant amplitude of
5 A and frequency of 8 kHz, which was applied between the lower contact (LC) and
the three contacts (RC) encircling the lower rim of the container (Fig. 2c). The resulting
current density $j_r$, which is concentrated in the bottom layer of the rubidium and basically
directed in radial direction, together with the strong vertical field $B_z$, generates an azimuthal
Lorentz force density $f_{\varphi} = j_r \times B_z$ that drives a torsional Alfv\'en wave in the 
fluid, as illustrated
in Fig. 2d. This upward propagating wave can be followed by the four stacked 
electric-potential probes PP 1-4, which measure the electromotive force (emf) $U = v_{\varphi} B_z r$ 
induced by
the azimuthal velocity component $v_{\varphi}$. As was shown in the supplemental material of 
\cite{Stefani2021}, the
correlation between the signals of those voltage probes allows us to identify the Alfv\'en speed
which depends on the instantaneous magnetic field $B_z(t)$ according to 
$v_a(t) = B_z(t)/(\mu_0 \rho)^{1/2}$ .
 
\begin{figure}[t]
  \centering
  \includegraphics[width=0.99\textwidth]{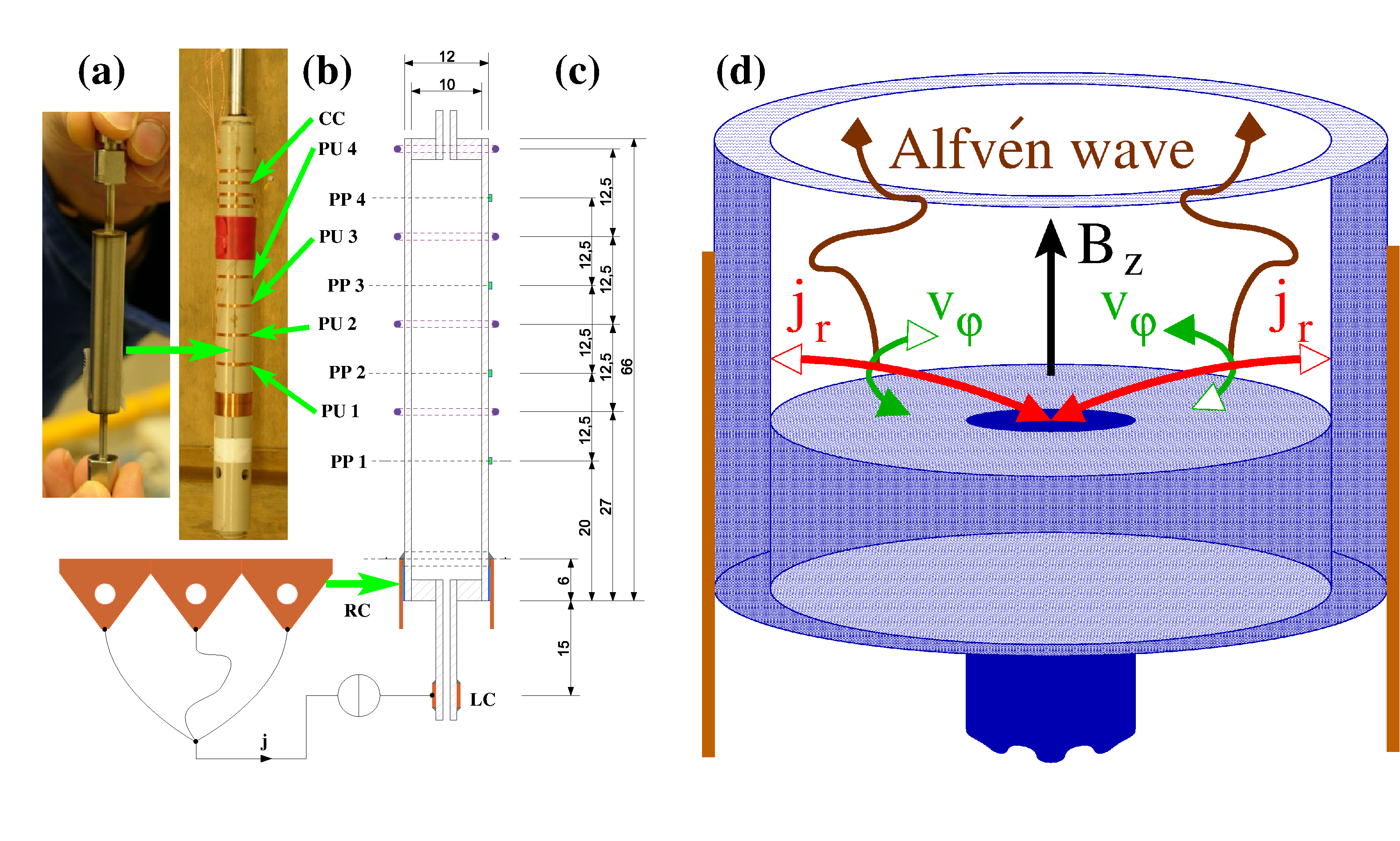}
  \caption{Experimental setting. (a) Stainless-steel container filled
with liquid rubidium. (b) Holder with four pickup coils (PU 1-4)
and four compensation coils (CC). (c) Geometrical details of the
construction. PP 1-4 denote four electric-potential probes
soldered on the container. The three orange triangles indicate the
rim contacts (RCs) encircling the bottom part of the container.
LC is the lower contact. (d) Schematic for the driving of the
torsional Alfv\'en wave in the lower part of the container. 
After \cite{Stefani2021}.
  }
  \label{fig:2}
\end{figure}

The four pick-up coils PU 1-4, in turn, measure the induction by the time-dependent
azimuthal current $j_{\varphi}$, which results from the interaction of radial velocity
components $v_r$ with $B_z$. The voltages measured at the pick-up coils are, though 
in a non-trivial
manner, representative of the sound wave in the liquid rubidium. As shown in 
\cite{Stefani2021}, their
spectrum contains a strong 16\,kHz component, arising from the ponderomotive force, 
but also
an 8\,kHz signal of comparable amplitude, an effect that has also been detected in 
a previous
experiment by Iwai \cite{Iwai2003}.

A first result of our numerical simulations 
of the nonlinear coupling
between the externally forced Alfv\'en waves
and the magneto-sonic with doubled frequency and vertical wavenumber
is shown in Fig. 3.  
The simulation was done with the compressible, finite-volume Godunov-type code 
PLUTO \cite{PLUTO}. The torsional Alfv\'en waves are continually driven by an external 
toroidal force, which is concentrated near the bottom of the cylinder and oscillates 
with a fixed frequency of 8 kHz, mimicking the Lorentz force in the experiment. The 
structure of resulting Alfv\'en waves consist of a large-scale standing wave with 
dominant toroidal velocity $v_{\varphi}$ and magnetic field perturbations $b_{\varphi}$, 
which vary in time with the same frequency as the forcing. Their vertical wavelength, 
determined by the Alfv\'en wave dispersion relation for this frequency, approximately 
fits into the container height (Fig. 3a, left panel). The magneto-sonic waves 
excited by the Alfv\'en waves have in turn a dominant vertical velocity, $v_z$, and 
magnetic field perturbations $b_z$. Their structure is shown in Fig. 3a, right panel, 
which consists of a standing wave with about twice shorter wavelength and doubled 
oscillating frequency (see Fig. 3b). 

\begin{figure}[t]
  \centering
  \includegraphics[width=0.99\textwidth]{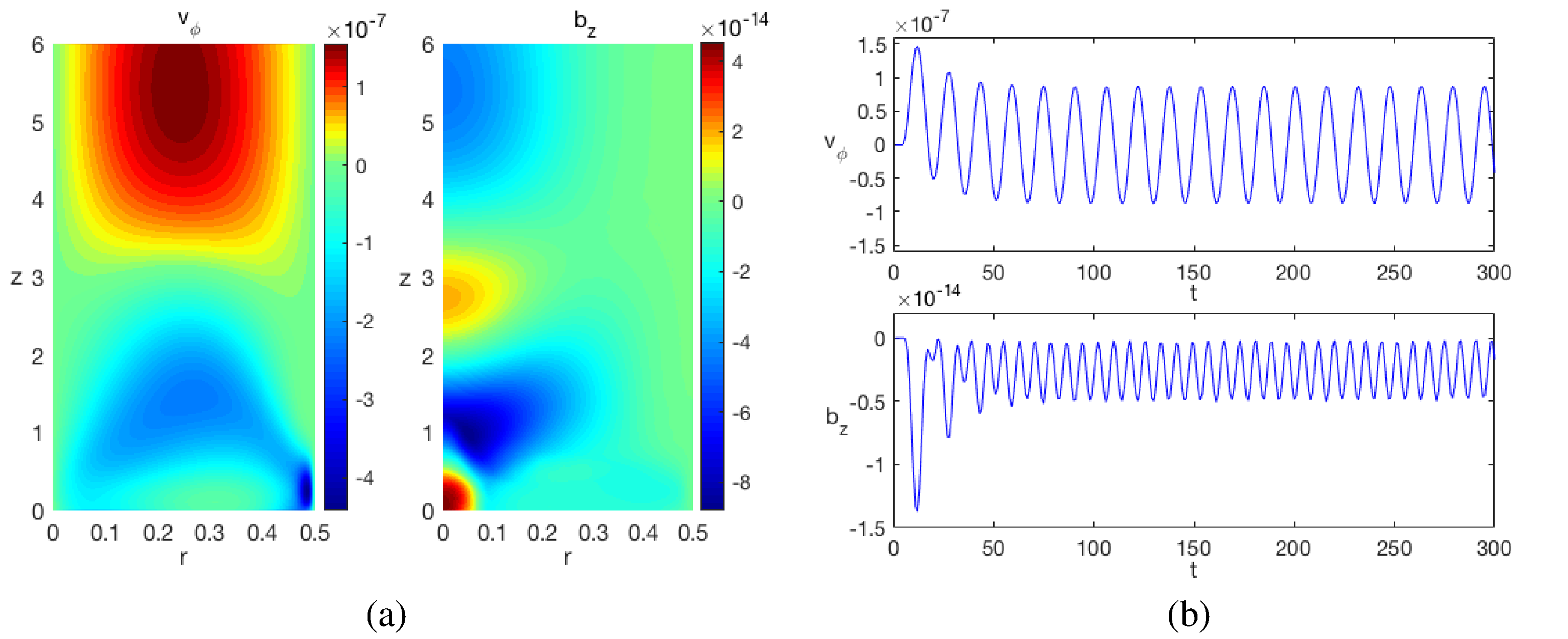}
  \caption{Numerical simulations of the Alfv\'en-wave experiment at a Lundquist number 
  of 45, corresponding to a magnetic field amplitude of  $B_0=35$\,T in the  experiment. 
  (a) Snapshots of the toroidal velocity $v_{\varphi}$ corresponding to the  Alfv\'en 
  wave driven by the external forcing, and the perturbation $b_z$ 
  of the vertical magnetic field 
  corresponding to the generated magneto-sonic waves in the $(r,z)$-plane. (b) 
  Time-development of these quantities at the fixed point $(r_0, z_0)=(0.25, 2.24)$ 
  (in units of the cylinder diameter $d=10$\,mm) in the container 
  (time in units of $d/c_s$). Note that the frequency 
  and wavenumber of the magneto-sonic waves are about twice larger than those 
  of the Alfv\'en waves.}\label{fig:3}
\end{figure}

\section{Main results}

Figure 4 summarizes the main results of one experimental run,
carried out with a temperature of 50$^\circ$C at which rubidium is liquid. The experiment starts by
charging the capacitor bank of HLD up to a voltage of 22\,kV. After releasing (at $t = 20$\,ms in
Fig. 4a) the stored energy, the axial magnetic field increases quickly to attain a maximum
value of 63.3\,T at $t = 53$\,ms. Afterwards, the field decays slowly, reaching a value of 2.1\,T at
the end of the interval shown here ($t = 150$\,ms). The period during which the field exceeds the
critical value of 54\,T extends from 40.5 to 66\,ms, as indicated by the dashed red lines
(see Fig. 4a).

\begin{figure}[t]
  \centering
  \includegraphics[width=0.99\textwidth]{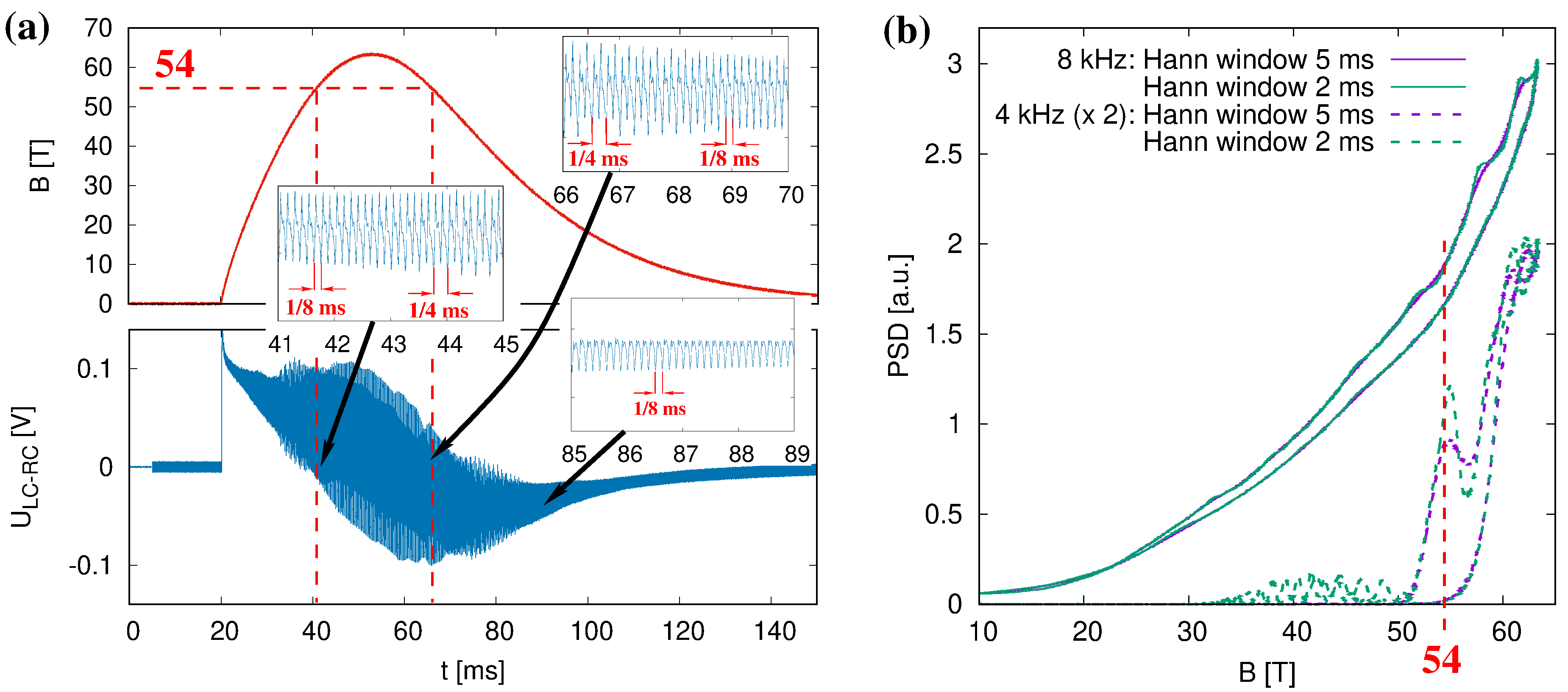}
  \caption{(a) Time dependence of the pulsed magnetic field and of
   the voltage measured at the lower
contact. The red dashed lines indicate the instants where the critical field 
strength of 54\,T is crossed. The insets of (a) detail one normal region in the 
falling branch (between 85 and 89 ms) with a mainly
pure 8 kHz signal, and the two transition regions, where the double-period signal due to torsional
Alfv\'en wave starts and ceases to exist. (b) Power spectral density (PSD) of the 8 and 4\,kHz 
coefficients of the voltage 
$U_{\rm LC-RC}$
from Figs. 4(a) and for two different von Hann windows. While the 8\,kHz signal shows a very
monotonic (quadratic) increase with the field, the 4 kHz signal appears only above the 
``magic point''
of 54 T. The peak at this very value, corresponding to the falling branch, confirms the theoretic
prediction of \cite{Zaqarashvili2006}. The increase of the signal beyond this field 
value is not yet fully understood. See \cite{Stefani2021}.
  }
  \label{fig:4}
\end{figure}

The voltage $U_{\rm LC-RC}$ measured between the contacts LC and RC comprises three
contributions: First, the usual Ohmic voltage drop over the contacts (with an amplitude of
approximately 5\,mV, as observed at the start and end of the magnetic-field pulse when the
magnetic field is close to zero); second, a certain long-term trend resulting from the time
derivative of the pulsed field; third, a significant electromotive force (EMF) 
$v_{\varphi} \times B_z$ arising
from the interaction of the toroidal velocity $v_{\varphi}$ of the generated 
torsional Alfv\'en wave with
the axial magnetic field $B_z$ .

Even without having a detailed numerical analysis at hand yet, we can plausibilize the measured
oscillation amplitude of this EMF: with an applied current amplitude of 5\,A, and taking into
account the contact geometry, we obtain a current 
density of about $j_r \approx 100$\,kA/m$ ^2$ , leading
(with an axial field $B_z = 50$\,T and a density of $\rho = 1490$\,kg/m$^3$ ) to an 
azimuthal acceleration of
$a_{\varphi} \approx 3000$\,m/s$^2$. Assumed to act over half an excitation 
period (i.e., 1/16\,ms), this acceleration
would generate flow velocities of $v_{\varphi} \approx 20$\,cm/s, 
which, when integrated over the
container's radius of $r = 5$\,mm), induce a voltage of $v_{\varphi} B_z r \approx 50$\,mV. 
Encouragingly, this very rough
estimate is in good agreement with the measured amplitude of the voltage oscillation as seen
in the high-field segment of Fig. 4a. In light of the high quality of the Alfv\'en-wave resonator
(as indicated by a Lundquist number of 67) this agreement is not that surprising.
The three insets of Fig. 4a show in more detail one ``normal'' segment within the falling
branch of the pulsed field (between 85 and 89\,ms), characterized by a more or less pure 8\,kHz
signal, and the two transition periods in the vicinity of 54\,T, where the double-period 4\,kHz
signal due to a secondary torsional Alfv\'en wave starts and ceases to exist.

The measured voltage $U_{\rm LC-RC}$ is now analyzed in detail by means of a windowed
Fourier transform (or Gabor transform), from which we can infer the instantaneous Power
Spectral Density (PSD) for different frequencies, including both the rising and falling
branches of the pulsed magnetic field. As shown in \cite{Stefani2021}, the dominant feature of this
spectrogram is, unsurprisingly, the 8\,kHz signal. More interesting, however, is the
emergence of a new signal at 4\,kHz at the threshold of 54\,T where $v_a = c_s$. Figure 4(b) confirms
the nearly quadratic dependence of the PSD of the externally driven 8\,kHz mode (with
only a slight difference between the rising and the falling branch), which one expects from the
linear dependence of the torsional velocity on $B_z$. We also observe a clear peak of the 
4\,kHz signal close to 54\,T, coming from the falling branch, and another peak behind, coming
from the rising branch. Actually, this persistence of the 4\,kHz signal for $v_a > c_s$ is not 
explained using the parametric resonance model of \cite{Zaqarashvili2006} without difficulty 
and,
therefore, needs a more detailed analysis.

\section{Conclusions}

In this paper, we have summarized the Alfv\'en-wave
experiments with liquid rubidium carried out at the HLD of HZDR which have, for the first
time, allowed us to cross the threshold $v_a = c_s$ where the transformation between magneto-sonic
and Alfv\'en waves becomes particularly efficient. Indeed, at that point we have observed a
secondary torsional wave with doubled period, in good agreement with the theoretical
prediction of swing-excitation between magneto-sonic and Alfv\'en waves. In a next step, we
will try to support the experimental work with more detailed numerical simulations using 
the PLUTO code. This will help us to better understand the observed effects, and also 
to optimize the parameters of
follow-up experiments.

\section*{Acknowledgments}
We acknowledge support of the HLD at HZDR, a member of
EMFL, and the DFG through the W\"urzburg-Dresden Cluster of Excellence on Complexity
and Topology in Quantum Matter--ct.qmat (EXC 2147, Project No. 390858490). F. S.
acknowledges further support by the European Research Council (ERC) under the European
Union's Horizon 2020 Research and Innovation Programme (Grant No. 787544). We thank
J\"urgen H\"uller for his assistance in the adventurous filling procedure of the rubidium
container, and Frank Arnold, Larysa Zviagina, Carsten Putzke, Karsten Schulz, and Marc
Uhlarz for their help in preparing and carrying out the experiment.

% ------------------------------------ Bib --------------------------------

%\bibliographystyle{pamir}
%\bibliography{references}

% -------------------- End of References

% ------------------------------------------------------------------------

\lastpageno	% This command sends the number of the last page to 
		% the PAMIR headline. Please latex your file twice if
		% you have used it.

%\clearpage

\end{document}